\documentclass[prl,floatfix,superscriptaddress,twocolumn]{revtex4}

\usepackage[dvips]{graphicx}

\def\av#1{\left\langle#1\right\rangle}
\usepackage{latexsym}
\usepackage{amssymb,amsfonts,amsmath}
\usepackage{epsfig}

\begin{document}

\title{The critical effect of dependency groups on the function of networks}

\author{Roni Parshani}
\affiliation{Minerva Center \& Department of Physics, Bar-Ilan University, Ramat Gan, Israel}
\author{Sergey V. Buldyrev} \affiliation{Center for Polymer
 Studies and Dept. of Physics, Boston Univ., Boston, MA 02215 USA}
\affiliation{Department of Physics, Yeshiva University, 500 West 185th
 Street, New York, New York 10033, USA}
\author{Shlomo Havlin}
\affiliation{Minerva Center \& Department of Physics, Bar-Ilan University, Ramat Gan, Israel}

\date{\today}

\begin{abstract}
Current network models assume one type of links to define the relations between the network entities.
However, many real networks can only be correctly described using two different types of relations. 
Connectivity links that enable the nodes to function cooperatively as a network and dependency links that bind the failure of one network element to the failure of other network elements.  
Here we present for the first time an analytical framework for studying the robustness of networks that include both connectivity and dependency links.
We show that the synergy between the two types of failures leads to an iterative process of cascading failures 
that has a devastating effect on the network stability and completely alters the known assumptions regarding the robustness of networks.
We present exact analytical results for the dramatic change in the network behavior when introducing dependency links.    
For a high density of dependency links the network disintegrates in a form of a first order phase transition while for a low density of dependency links the network disintegrates in a second order transition. 
Moreover, opposed to networks containing only connectivity links where a broader degree distribution results in a more robust network, 
when both types of links are present a broad degree distribution leads to higher vulnerability.
\end{abstract}
\maketitle

Many friendships between individuals in a social network, numerous business connections in a financial network 
or multiple cables between Internet routers, are all examples of networks with a high density of connectivity links \cite{ws,barasci,bararev,PastorXX,mendes,Barrat,referee1,referee2,Newman-press,mnewman}. 
Such networks are regarded as very stable to attacks since even after a failure of many nodes the network still remains connected.
In contrast, dependencies between the network nodes endanger the network stability since the failure of several 
nodes may lead to the immediate failure of many others.
As an example consider a financial network: Each company has trading and sales connections with other companies (connectivity links).
These connections enable the companies to interact with each other and function together as a global financial market. But there are also dependencies relations between companies, several companies that belong to the same owner depend on one another. 
If one company fails the owner might not be able to finance the other companies that will fail too. 
Such dependencies jeopardize the network stability and are the possible cause of many major financial crises. 
Another example is an online social network (Facebook or Twitter): Each individual communicates with his friends (connectivity links), thus forming a 
social network through which information and rumors can spread. 
However, many individuals will only participate in a social network if other individuals
with common interests also participate (dependency links) in that social network. 

\begin{figure}[h]
\begin{center}
\epsfig{file=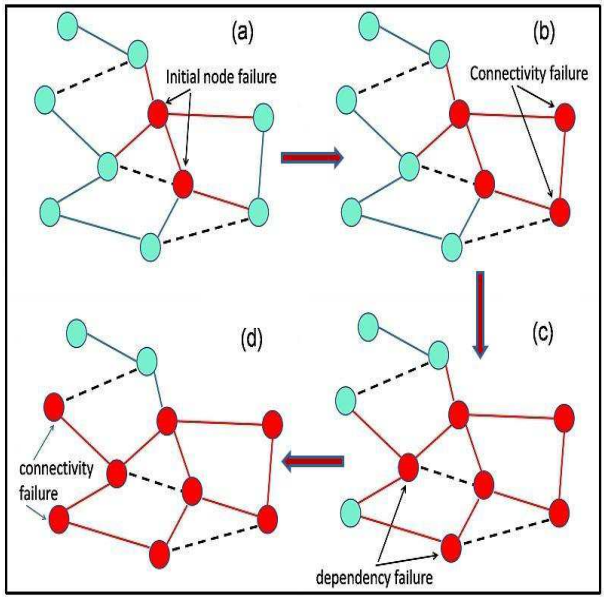,height=8cm,width=8cm}
\end{center}
\caption{Demonstration of the synergy between the percolation process 
and the failures caused by dependency links (dependency process) that lead to an iterative process of cascading failures. 
The network contains two types of links: connectivity links (solid lines) and dependency links (dashed lines).
(a) The process starts with the initial failure of two nodes (marked in red). The connectivity links connected to them also fail (marked in red). 
(b) Percolation process - in this stage all the nodes and the connectivity links that are connected to them, that are not connected to the giant    
    cluster (largest cluster) by connectivity links also fail (marked in red). 
(c) Dependency process - the nodes that depend (connected by dependency links) on the failing nodes also fail (marked in red).
(d) The next step of connectivity failure in which two more nodes fail because they are not connected to the largest 
    cluster (currently containing only two nodes).} 
\label{demonstration}
\end{figure}

The effect of failing nodes on the network stability has been studied separately for networks containing only 
connectivity links \cite{ER,ER2,Bollobas,cohena,Callaway,HavlinBook} and for networks containing only dependency links \cite{motter_cascade,motter2_cascade,power_cascade,transportation_cascade,agent_cascade}.
The fundamental difference between connectivity and dependency links is that for dependency links the failure of a direct neighbor of a node  leads to the direct failure (with some probability) of that node, but for connectivity links a node fails only when it (or the cluster it is in) becomes completely disconnected from the network. Percolation theory is a major tool for studying network stability when the network is connected only with connectivity links. In a percolation process on a network of size $N$, a fraction $1-p$ of the network nodes are removed. 
If the remaining fraction of nodes, $p$, is larger then a critical value ($p>p_c$), a spanning cluster connecting order $N$ nodes exists,
if however, $p<p_c$, the network collapses into small clusters. At $p=p_c$ the network undergoes a second order phase transition \cite{ER,ER2,Bollobas,cohena,Callaway,HavlinBook}. 

Previous studies of networks containing dependencies can be divided into two categories: 
(i) Overload failures in networks containing a flow of a physical quantity. 
For example, disturbances in power transmission systems or congestion instabilities in transportation 
networks and Internet traffic \cite{motter_cascade,motter2_cascade,power_cascade,transportation_cascade}.
These models show that when one node is overloaded and the traffic cannot be routed through it, 
choosing alternative paths will cause other nodes to also become overloaded. 
This process may develop into a series of cascading failures that can disable the entire network. 
(ii) Models based on local dependencies, such as decision making of interacting agents \cite{agent_cascade}. 
In these models the state of a node depends on the state of its neighbors and therefore a failing node 
will cause it's neighbors to also fail and so on. 
\begin{figure}[h]
\begin{center}
\epsfig{file=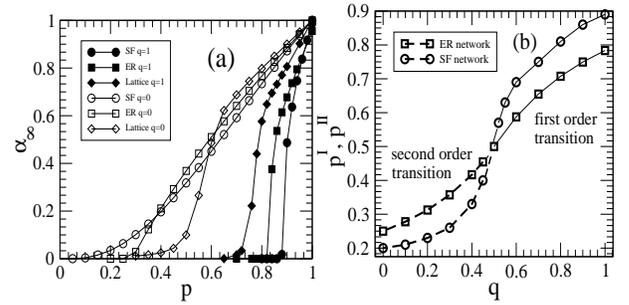,height=4cm,width=8cm}
\caption{ (a) Simulation results showing the first and second order phase transitions in lattice, ER and SF networks. 
The fraction of nodes in the giant component at the end the cascade process, $\alpha_{\infty}$, is shown as a function of $p$ 
for $q=1$ (filled symbols) and for $q=0$ (open symbols), where $q$ is the fraction of dependent nodes. 
For $q=1$, $\alpha_{\infty}$ abruptly drops to zero at the transition point characterizing a first order transition.      
For $q=0$, $\alpha_{\infty}$ gradually approaches zero as expected in a second order transition.
The SF (circle) and ER (square) networks presented both have the same average degree of $\av{k}=3.5$. Thus, SF networks that are most robust 
when only connectivity links exist (very low transition point for $q=0$) become most vulnerable when dependency links 
are added (very high transition point for $q=1$).
(b) The transition points, $p^I$ for the first order region (solid line) and $p^{II}$ for the second order region (dashed line) are plotted as a function of $q$ (the fraction of dependent nodes) for ER (squares) and SF (circles) networks with the same average degree $\av{k}=4$. 
For ER networks theoretical results (confirmed by simulation results) are obtained according to Eq.(\ref{eq_cond1}) and Eq.(\ref{eq_cond2}) 
presented in the paper. 
For SF networks simulation results of a network with $\lambda=2.9$ are presented since exact theoretical results are not available.} 
\label{firstSecondTransition}
\end{center}
\end{figure}
\section{Results}
Here we present an analytical framework for studying the robustness of networks that include both connectivity and dependency links.
When nodes fail in a network containing both types of links, two different processes occur. 
(i) Connectivity links are disconnected, causing other nodes to disconnect from the network (percolation process). 
(ii) Failing nodes cause other nodes that depend on them to also fail even though they are still connected via connectivity links (dependency process). 
We show that the synergy between the percolation process and the dependency process leads to a cascade of failures that can fragment 
the entire network (Fig.\ref{demonstration}).
We find that the density of dependency links, $q$, plays a key role in determining the robustness of such networks.   
For networks containing connectivity links and a high density of dependency links, 
an initial failure of even a small fraction of the network nodes disintegrates the network in a form of a first order phase transition. 

If however, the fraction of dependency links is reduced below a certain threshold, $q_c$, the network disintegrates 
in a form of a second order phase transition. The cascading process leading to a first order transition exists for a wide range of topologies 
including lattices, ER and SF networks, indicating it is a general property of many networks (Fig.\ref{firstSecondTransition}(a)).
Comparing networks with both connectivity and dependency links but with different topologies, 
reveals a new relation between topology and the robustness to random failure: 
Networks with a broader degree distribution of connectivity links are more vulnerable to random failure in the presence of dependency links. 
This is opposed to the known result for networks containing only connectivity links, 
where networks with a broader degree distribution are significantly more robust to random failures.  
Fig.\ref{firstSecondTransition}(a) and Fig.\ref{firstSecondTransition}(b) show that when comparing ER and SF networks 
with the same average degree, SF networks with a high density of dependent nodes are more vulnerable to random failures then ER networks. 
 
\section{Formalism}
Next we present an analytical approach for studying the robustness to random failure of networks containing the two types of links. 
Without loss of generality we define a model in which only pairs of nodes depend on one another, forming dependency groups of size 2. 
When the dependency group contain more then two nodes the cascade effect is even more extreme and the transition from the regular second order percolation transition to a first order transition occurs even for more stable networks (see SI). Therefore, the new properties we present for the case of dependency groups of size 2 are also valid in the general case of larger dependency groups (see Fig.1 in SI).    
The model is defined as follows: A network containing $N$ nodes is randomly connected by connectivity links 
with a degree distribution $P(k)$ and an average degree $\av{k}$. In addition, pairs of nodes are connected by dependency 
links as follows: 
a) A node can only have one dependency link. b) If node $i$ depends on node $j$ then node $j$ 
depends on node $i$. For this model we denote by $q$ the fraction of nodes that have dependencies.                    

We start by presenting the formalism describing the iterative process of cascading failures for the simple case of $q=1$ (see Fig.1 in SI).
Each iteration (step) includes failures that are the result of the percolation process and failures that are the result of the dependency process. The goal of the formalism is to describe the accumulated process up to step $n$ 
as an equivalent single random removal, $r_n$, from the original network.
The remaining fraction of nodes after such a removal is $\beta_n = 1- r_n$.
The new network after the removal of a fraction $r_n$ of the nodes, 
has a giant component consisting of a fraction $g(\beta_n)$ of the remaining nodes which is a fraction $\alpha_{n+1}=\beta_{n}g(\beta_{n})$ 
from the original network.\\
The iterative process  is initiated by the removal of a fraction $r_0=1-p$ of the network nodes. The remaining part of the network is $\beta_0=p$.
This initial removal will cause additional nodes to disconnect from the giant cluster due to the percolation process.
The fraction of nodes that remain functional after the percolation process is $\alpha_1= \beta_0 g(\beta_0)$.
Each node from the non functional part ($1-\alpha_1$) will cause the node that depends on it to also fail (dependency process).
The probability that a node depending on a non functional nodes has survived until now is $\alpha_1$. 
Therefore the fraction of new nodes that will fail due to dependencies is $\delta_1=(1-\alpha_1)\alpha_1$.
The accumulated failure including the initial failure of $1-\beta_0$ and $\delta_1$ is equivalent to a random removal of $r_1 = (1-\beta_0) + (1-\alpha_1)\beta_0$ from the original network (see SI).
The remaining fraction of nodes after the new removal is therefore $\beta_1=1-r_1=\beta_0\alpha_1=\beta_0^2g(\beta_0)$. 
The remaining functional part of the giant component is now $\alpha_2=\beta_1 g(\beta_1)$.
To calculate the fraction  $\delta_2$ of nodes that are disconnected due to dependencies at the second stage, 
recall that at the previous stage a fraction $\delta_1$ failed from $\alpha_1$.
The remaining part of $\alpha_1$ was therefore $\alpha_1 - \delta_1 = \alpha_1^2$.
Thus $\delta_2=(\alpha_2/\alpha_1^2)(\alpha_1^2-\alpha_2)=[1-(\alpha_2/\alpha_1^2)]\alpha_2$.
This is equivalent to a random removal of $r_2 = (1-\beta_1) + [1-(\alpha_2/\alpha_1^2)]\beta_1$ from the original network.
The remaining fraction of nodes is $\beta_2=1-r_2=\alpha_2\alpha_1^2/\beta_1=\beta_0^2g(\beta_1)$.
Following this approach we can construct the sequence, $\beta_n$, of the remaining fraction of nodes in the network after each iteration.\\
$\beta_0=p$.\\
$\beta_1=p^2g(\beta_0)$.\\
$\beta_2=p^2g(\beta_1)$...\\
$\beta_n=p^2g(\beta_{n-1})$.\\

\begin{figure}[h]
\begin{center}
\epsfig{file=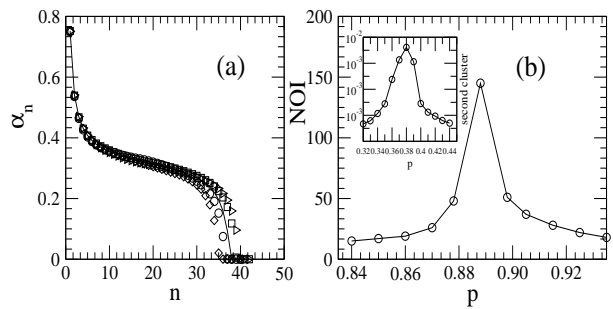,height=4cm,width=8cm}
\end{center}
\caption{ (a) Comparison between simulations and theoretical results for the fraction of nodes in the giant cluster on every step $n$ of the iterative process of failing nodes. The results are shown for an ER network with $q=0.8$ and $p=0.84$ ($p \simeq p^{I}$). 
The theoretical results (line) are calculated according to Eq.(\ref{eq1}) (the explicit form of $g(x)$ is presented in the text) 
and are compared to several realizations of computer simulations on networks of size $N=200K$. 
(b) The number of iterative failures (NOI) are shown for a scale free network with $\lambda=2.7$ and $q=1$. 
At the first order transition point, the number of iterative failures that the network undergoes before disintegrating scales as $N^{1/4}$ (see SI).
This number sharply drops as the distance from the transition is increased. Thus, plotting the number of iterations as a function of $p$ provides a useful method for identifying the transition point, $p^{I}$, at the first order region. The inset shows that the size of the second largest cluster reaches its maximum value at the second order transition point, $p^{II}$, therefore providing a useful method for identifying $p^{II}$ at the second order region.} 
\label{iterationsOfGiantComponent}
\end{figure}

\noindent Following a similar approach for the general case of $0 \leq q \leq  1$ (see SI) yields the sequence
$\beta_n=qp^2g(\beta_{n-1}) + p(1-q)$. Given, $\beta_n$, the fraction of nodes in the giant cluster is $\alpha_{n+1}=\beta_{n}g(\beta_{n})=p(1-q(1-pg(\beta_n)))g(\beta_n)$.
Fig.~\ref{iterationsOfGiantComponent}(a) compares theory and simulations of $\alpha_n$, for the case of an ER network. 

\noindent To determine the state of the system at the end of the cascade process we analyze $\beta_{n}$ at the limit of $n \to \infty$.
This limit must satisfy the equation $\beta_{n}$=$\beta_{n+1}$ since at the end of the process the cluster is not further fragmented.
Denoting $\beta_n=\beta_{n-1}=x$ we arrive to the equation:
\begin{equation}
x=p^2qg(x)+p(1-q);
\label{eq1}
\end{equation}
This equation can be solved graphically as the
intersection of a straight line $y=x$ and a curve $y=p^2qg(x)+p(1-q)$. 
When $p$ is small enough the curve increases very slowly and does not 
intersect with the straight line (except at the origin which corresponds to the trivial solution). 
The critical case for which the nontrivial solution emerges,
corresponds to the case when the line touches the curve at a single
point $x$ and in this point we have the condition $1=p^2q \frac{d g}{dx}(x)$, which together with
Eq.(\ref{eq1}) gives the solution for the critical fraction of failing nodes that will fragment the network and the
critical size of the giant component.

\section{Analytical solution}
An exact analytical solution can be obtained using the apparatus of
generating functions. As in Refs.~\cite{Newman,Shao,Buldyrev} we introduce the generating function of the
degree distribution $G_{0}(\xi)=\sum_k P(k) \xi^k$.
Analogously, we also introduce the generating function of the underlying
branching process, $G_{1}(\xi)=G'_{0}(\xi)/ G'_{0}(1)$.
A random removal of a fraction $1-p$ of nodes will change the degree
distribution of the remaining nodes, so the generating
function of the new distribution is equal to the generating function of
the original distribution with the argument $\xi$ replaced by $1-p(1-\xi)$ \cite{Newman}. 
The fraction of nodes that belong to the giant component after the removal of $1-p$ nodes is $g(p)=1-G_{0}[1-p(1-f)]$, 
where $f=f(p)$ satisfies a transcendental equation $f=G_{1}[1-p(1-f)]$ \cite{Shao}.

In the case of an ER network with a Poisson degree distribution \cite{ER,ER2,Bollobas}, the
problem can be solved explicitly since $G_{1}(\xi)=G_{0}(\xi)=\exp(\av{k}(\xi-1))$. 
Accordingly, $g(x)=1-f$ and $f=\exp[\av{k}x(f-1)]$ where $x$ is defined in Eq.(\ref{eq1}).
The fraction of nodes in the giant component at the end of the cascade process is then given by $\alpha_\infty=\beta_\infty g(\beta_\infty)=p(1-q(1-p(1-f)))(1-f)$.
The equation $f=f(q,p,k)$ has a trivial solution at $f=1$. The non-trivial solutions of $f$ can be presented by the crossing points of the two curves in a system of equations that are given with respect to $x$ and $f$:
\begin{equation}
\left\{
\begin{array}{lr}
x=p^2q(1-f)+p(1-q)\\
x=\frac{\ln{f}}{\av{k}(f-1)}.
\end{array}
\hspace{1.5cm} , 0 \leq f < 1
\right .\
\label{system0}
\end{equation}
%
For the trivial solution at $f=1$ the size of the giant component is zero ($\alpha_\infty=0$).
For the solutions that are the crossing points of the two curves, $f < 1$, i.e., $\alpha_\infty>0$.
Thus, the case where the curves tangentially intersect corresponds to a {\it first order} phase transition point ($p = p^I$) 
where $\alpha_\infty$ abruptly jumps from a finite size above $p^I$ to zero below $p^I$ \cite{Raissa}.
The condition for the first order transition is that the derivatives of the equations of system (\ref{system0}) with respect to $f$ are equal.
Together with system (\ref{system0}) this yields: 
\begin{equation}
{(p^I)}^2\av{k}q=-1/[(f-1)f]+\ln f/(f-1)^2
\label{eq_cond1}
\end{equation}
However, for a solution of system (\ref{system0}) where $f \to 1$ ($\alpha_\infty=0$) there is no jump in the size of the giant cluster and thus 
the transition is a {\it second order} transition ($p=p^{II}$). Solving system (\ref{system0}) for $f \to 1$ yields:\\
\begin{equation}
p^{II}\av{k}(1-q)=1
\label{eq_cond2}
\end{equation}
The analysis of Eq.(\ref{eq_cond1}) and Eq.(\ref{eq_cond2}) shows that the first order transition at $p=p^I$ occurs for networks with a high density of dependency links ($q>q_c$), 
while the second order transition at $p=p^{II}$, occurs for networks with a low density of dependency links ($q<q_c$).
This is confirmed by Fig. 4(a) that compares theory and simulations for $p^{II}(q)$ and  $p^{I}$(q).
The critical value of $q_c$ (and $p_c$) for which the phase transition changes from first order to a second order is 
obtained when the conditions for both the first and second order transitions are satisfied simultaneously. Applying both conditions we obtain  

\begin{equation}
\left\{
\begin{array}{lr}
q_c=(\av{k}+1-\sqrt{2\av{k}+1})/\av{k}\\
p_c=1/(\sqrt{2\av{k}+1}-1).\\
\end{array}
\right.\
\label{system1}
\end{equation}

\section{Simulations}
Next, we support our analytical results by simulations. Finding the transition point via simulations is 
always a difficult task that requires high precision. In the case of the first order transition we are able to calculate the transition point with good precision by identifying the special behavior characterizing the number of iterations (NOI) in the cascading process. 
\begin{figure}[h]
\begin{center}
\epsfig{file=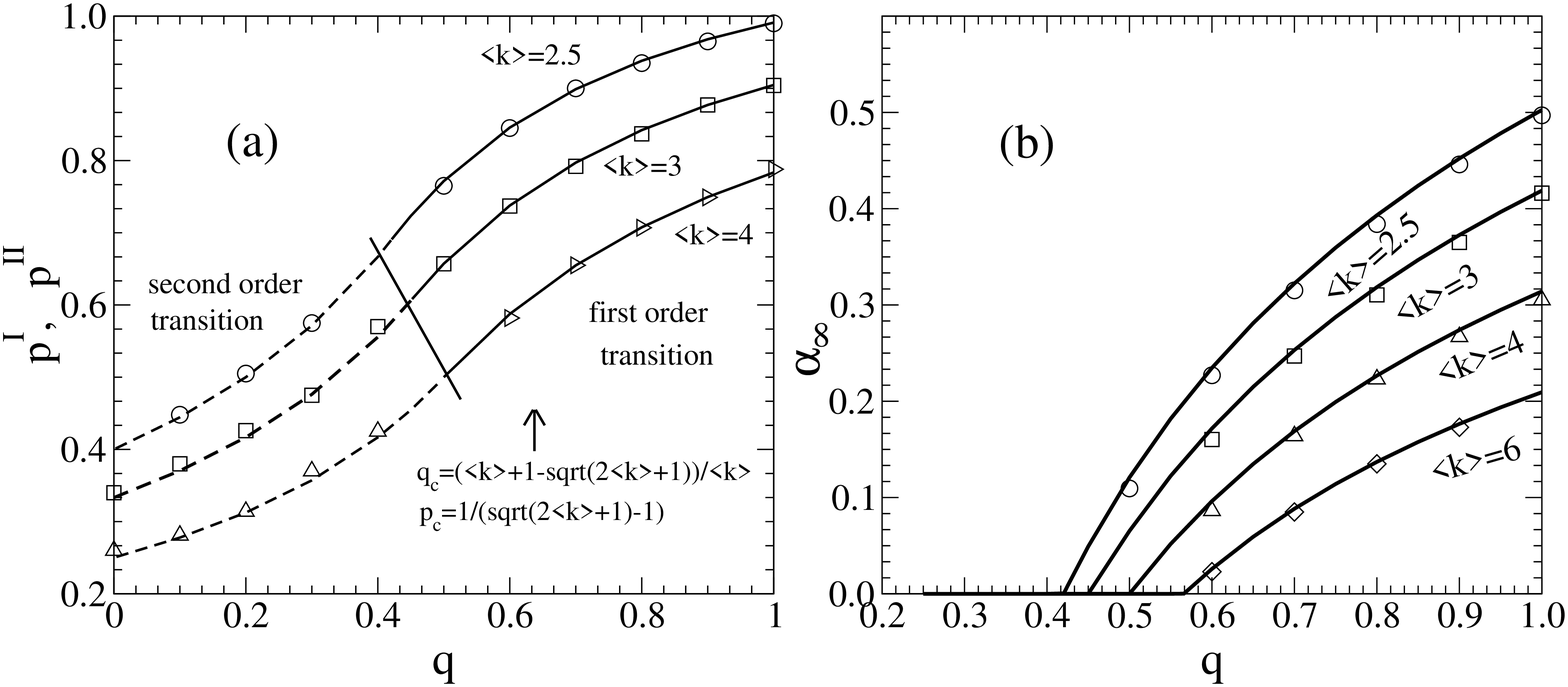,height=4cm,width=8cm}
\end{center}
\caption{ (a) Theory (lines) and simulations (symbols) are compared for the values of $p^I(q)$ and $p^{II}(q)$ for ER networks with different values of $\av{k}$. 
For $q > q_c$ the network undergoes a first order transition, therefore the theoretical values of the transition point, $p^I(q)$, that are calculated according to Eq.(3) are compared with simulations performed using the NOI method (explained in text).
For $q < q_c$ the network undergoes a second order transition, therefore the theoretical values of the transition point, $p^{II}(q)$, that are calculated according to Eq.(4) are compared with simulations performed using the second largest cluster method (explained in text).
The line separating between the first and second order is obtained according to Eq.(5).   
(b) Comparison between simulation (symbols) and theory (lines) for $\alpha_{\infty}$ as a function $q$ for different values of $\av{k}$. 
$\alpha_{\infty}$, at the phase transition point is finite for a first order transition and a zero fraction for a second order transition.} 
\label{fff}
\end{figure}
At the first order transition point, the NOI scales as $N^{1/4}$ (see SI) which is also demonstrated by the long plateau 
in Fig.~\ref{iterationsOfGiantComponent}(a). This number sharply drops as the distance from the transition point is increased, since away from 
the transition point, $p^I$, the NOI scales as $\log{N}/(p-p^I)$ (see SI). 
Thus, plotting the NOI as a function of $p$, provides a useful and precise method for identifying the transition point $p^I$ at the first order region.
For the second order region a similar behavior exists for the size of the second largest cluster which also
reaches its maximum at the transition point \cite{HavlinBook}. Fig.~\ref{iterationsOfGiantComponent}(b) presents simulation results of the NOI. 
The transition point, $p^I$, can easily be identified by the sharp peek characterizing the transition point. 
The inset of Fig.~\ref{iterationsOfGiantComponent}(b) presents a similar behavior for the size of the second largest cluster near the second order transition point, $p^{II}$. Fig. 4(a) compares simulation results and theory for the transition points $p^I(q)$ at the first order region (solid line) and $p^{II}(q)$ at the second order region (dashed line). 
The transition points were obtained using the NOI and the second cluster size techniques respectively. 
The theoretical results for different values of $q$ and $\av{k}$ were calculated by solving system (\ref{system0}) together with Eq.(\ref{eq_cond1}) or Eq.(\ref{eq_cond2}) respectively. 
Fig.~\ref{firstSecondTransition}(b) compares the values of the transition points $p^I(q)$ and $p^{II}(q)$ respectively between SF and ER networks with the same average degree. 
For networks with a small fraction of dependencies (second order transition region) SF networks are more robust to random failure (lower $p^{II}$). 
For networks with a high fraction of dependencies (first order transition region) SF networks become more vulnerable (higher $p^{I}$).
Fig. 4(b) compares simulation and theory for $\alpha_{\infty}$, the fraction of nodes in the giant cluster at the transition point. Above $q_c$, $\alpha_{\infty}$ is finite characterizing a first order transition, while below $q_c$, $\alpha_{\infty}$ is zero as expected for a second order transition.

\section{Discussion}

Here we show that in order to properly model real networks two different type of links are needed: connectivity links and dependency links.
We present an analytical formalism for a general network model including both connectivity and dependency links.
According to our model, networks with high density of dependency links are extremely vulnerable to random failure and when a critical fraction of nodes fail the network disintegrates in a form of a first order phase transition.
Networks with a low density of dependency links are significantly more robust and disintegrate in a form of a second order phase transition. 
In the limit of zero fraction of dependency links our general solution yields the known results for networks with only one type of links.
Our framework also provides an analytical solution for the critical density of dependency links for which the phase transition changes from a first order to a second order percolation transition. We develop a powerful simulation method for accurately estimate the transition point, based on the unique behavior of the NOI (number of iterations in the iterative process of cascading failures) that diverges at the first order transition point. Using this method we are able to provide very accurate simulation results supporting our analytical results.

\begin{acknowledgments}
We thanks the European EPIWORK project, the Israel Science Foundation, the ONR and the DTRA for financial support.
S.V.B. thanks the Office of the Academic Affairs of Yeshiva University for funding the Yeshiva
University high-performance computer cluster and acknowledges the
partial support of this research through the Dr. Bernard W. Gamson
Computational Science Center at Yeshiva College.
\end{acknowledgments}

\end{document}